\documentclass[3p,times,procedia]{elsarticle}
\usepackage{nupha_ecrc}

\volume{00}

\firstpage{1}

\journalname{Nuclear Physics A}

\runauth{}

\jid{nupha}

\jnltitlelogo{Nuclear Physics A}

\usepackage{amssymb}
\usepackage[figuresright]{rotating}

\begin{document}

\begin{frontmatter}

%% Title, authors and addresses

%% use the tnoteref command within \title for footnotes;
%% use the tnotetext command for the associated footnote;
%% use the fnref command within \author or \address for footnotes;
%% use the fntext command for the associated footnote;
%% use the corref command within \author for corresponding author footnotes;
%% use the cortext command for the associated footnote;
%% use the ead command for the email address,
%% and the form \ead[url] for the home page:
%%
%% \title{Title\tnoteref{label1}}
%% \tnotetext[label1]{}
%% \author{Name\corref{cor1}\fnref{label2}}
%% \ead{email address}
%% \ead[url]{home page}
%% \fntext[label2]{}
%% \cortext[cor1]{}
%% \address{Address\fnref{label3}}
%% \fntext[label3]{}

\dochead{}
%% Use \dochead if there is an article header, e.g. \dochead{Short communication}
%% \dochead can also be used to include a conference title, if directed by the editors
%% e.g. \dochead{17th International Conference on Dynamical Processes in Excited States of Solids}

\title{The unreasonable effectiveness of hydrodynamics in heavy ion collisions}

%% use optional labels to link authors explicitly to addresses:
%% \author[label1,label2]{<author name>}
%% \address[label1]{<address>}
%% \address[label2]{<address>}

\author{Jacquelyn Noronha-Hostler$^{1,2}$, Jorge Noronha$^{3}$, and Miklos Gyulassy$^{1}$}

\address{$^{1}$Department of Physics, Columbia University, 538 West 120th Street, New York,
NY 10027, USA \\

$^{2}$Department of Physics, University of Houston, Houston, TX 77204, USA \\

$^{3}$Instituto de F\'{\i}sica, Universidade de S\~{a}o Paulo, C.P. 66318,
05315-970 S\~{a}o Paulo, SP, Brazil}

\begin{abstract}
Event-by-event hydrodynamic simulations of AA and pA collisions involve initial energy densities with large spatial gradients. This is associated with the presence of large Knudsen numbers ($K_n\approx 1$) at early times, which may lead one to question the validity of the hydrodynamic approach in these rapidly evolving, largely inhomogeneous systems. A new procedure to smooth out the initial energy densities is employed to show that the initial spatial eccentricities, $\varepsilon_n$, are remarkably robust with respect to variations in the underlying scale of initial energy density spatial gradients, $\lambda$. For $\sqrt{s_{NN}}=2.76$ TeV LHC initial conditions generated by the MCKLN code, $\varepsilon_n$ (across centralities) remains nearly constant if the fluctuation scale varies by an order of magnitude, i.e., when $\lambda$ varies from 0.1 to 1 fm. Given that the local Knudsen number $K_n\approx \frac{1}{\lambda}$, the robustness of the initial eccentricities with respect to changes in the fluctuation scale suggests that the vn's cannot be used to distinguish between events with large $K_n$ from events where $K_n$ is in the hydrodynamic regime. We use the 2+1 Lagrangian hydrodynamic code v-USPhydro to show that this is indeed the case: anisotropic flow coefficients computed within event-by-event viscous hydrodynamics are only sensitive to long wavelength scales of order $\frac{1}{\Lambda_{QCD}}\approx 1$ fm and are incredibly robust with respect to variations in the initial local Knudsen number. This robustness can be used to justify the somewhat unreasonable effectiveness of the nearly perfect fluid paradigm in heavy ion collisions.
\end{abstract}

\begin{keyword}
Relativistic hydrodynamics \sep smoothing scale \sep anisotropic flow harmonics
%% keywords here, in the form: keyword \sep keyword

%% MSC codes here, in the form: \MSC code \sep code
%% or \MSC[2008] code \sep code (2000 is the default)
\end{keyword}

\end{frontmatter}

\section{Introduction}

\begin{figure}[htb]
\centering
\includegraphics[height=2.5in]{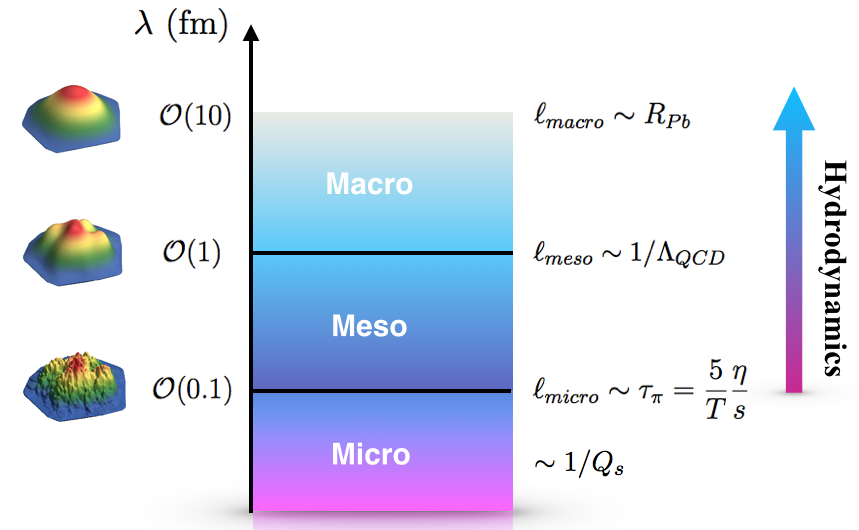}
\caption{ (Color online)  Picture of the various relevant energy scales within heavy ion collisions according to the smoothing scale $\lambda$.   The energy density snapshots on the right illustrate how a given initial condition changes with the
smoothing scale.}
\label{fig:scales}
\end{figure}

\begin{figure}[htb]
\centering
\includegraphics[height=1.6in]{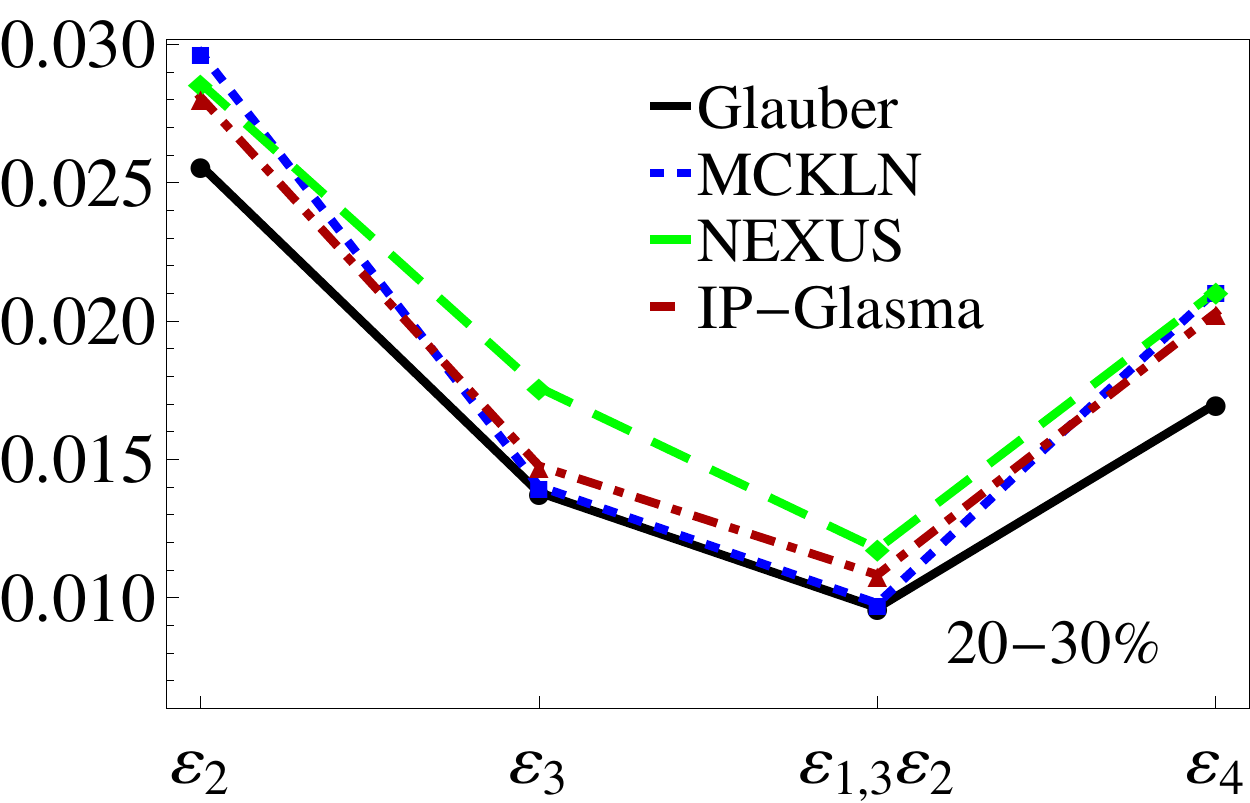} 
\caption{(Color online) Various eccentricities $\varepsilon_{n,m}$ for NeXus \cite{Drescher:2000ec,Gardim:2012yp}, Glauber \cite{Drescher:2006ca}, MCKLN \cite{Drescher:2006ca}, and IP-Glasma initial conditions \cite{Schenke:2012wb} at $20-30\%$ centrality.}
\label{fig:ec}
\end{figure}

A lingering question in the field of heavy-ion collisions is what energy scales are probed with different types of experimental observables.  As in the the early universe where inhomogeneities led to the formation of large scale structure after expansion \cite{weinbergcosmo}, the initial conditions for the Quark Gluon Plasma exhibit energy density fluctuations that originate from different physical scales as shown in Fig. \ref{fig:scales}.   The largest scale is the macroscale that is defined by the system size itself on the order of about $\approx 7-10$ fm for a $PbPb$ collision at LHC, which is encapsulated in the eccentricities of the initial collision \cite{Noronha-Hostler:2015coa}.  The upper limit of the smallest scale is the microscale, which in current viscous hydrodynamic simulations is defined by the relaxation time \cite{Denicol:2011fa} the smallest value of which is around $\tau_{\pi}\approx 0.25$ fm at LHC (the gluon saturation scale is even shorter $\approx 1/Q_s <0.1$ fm).  Between the microscale and the macroscale lies the mesoscale on the order of the size or a proton $\approx 1$ fm, which one would expect to be the defining scale for Glauberian dynamics where the degrees of freedom are only that of the nucleons.  It is well-known that initial conditions that include gluon saturation contain non-linear dynamics that produce different initial eccentricities compared to Glauberian initial conditions as seen in Fig.\ \ref{fig:ec}. However, comparing Glauber \cite{Miller:2007ri} vs. IP-Glasma \cite{Schenke:2012wb} initial conditions also produces a very different scale of energy density fluctuations.

Different types of initial conditions probe different scales in Fig. \ref{fig:scales}.  Smoothed, single shot initial conditions with no event-by-event fluctuations would have the largest $\lambda$ at the macroscale level.  Event-by-event calculations using Glauberian wounded nucleon dynamics would be at the mesoscale level whereas IP-Glasma initial conditions that include sub-nucleonic fluctuations are at the microscale level.  In order to explore the difference between these types of initial conditions, a cubic spline smoothing function is employed as discussed in  \cite{Noronha-Hostler:2015coa} to smooth out the initial conditions to each different scale. With this procedure the eccentricities are found to remain nearly constant when $\lambda \lesssim 1$ fm. Using the cubic spline all fluctuations underneath the scale $\lambda$ are smoothed out.  It is important that with this procedure the eccentricities remain nearly constant because they are the dominant force when it comes to determining the final flow harmonics \cite{Gardim:2011xv,Gardim:2014tya}. In the future it would be interesting to study the effects of the smoothing scale not only on the initial energy density but on the full initial energy-momentum tensor.

All calculations are done using MCKLN initial conditions \cite{Drescher:2006ca} within the event-by-event relativistic viscous hydrodynamical code v-USPhydro  \cite{Noronha-Hostler:2013gga,Noronha-Hostler:2014dqa} that is based upon the Lagrangian method to solve the equations of motion known as Smoothed Particle Hydrodynamics \cite{Aguiar:2000hw}.  Decays are modeled through an adapted version of the AZHYDRO code \cite{azhydro} with masses up to $M=1.7$ GeV although it would be interesting to test the influence of heavier resonances \cite{Noronha-Hostler:2013rcw}, which would likely further obscure effects from initial energy density fluctuations \cite{NoronhaHostler:2007jf,NoronhaHostler:2009cf,Noronha-Hostler:2014usa}.  The lattice based equation of state from \cite{Huovinen:2009yb} is employed.  Relativistic hydrodynamics is switched on at $\tau_0=0.6$ fm and the freeze-out temperature is varied between $T_{FO}=120-130$ MeV.  The shear viscosity to entropy density ratio is held constant such that the best fitting $\eta/s$ is used for each individual $\lambda$, although a temperature dependent $\eta/s(T)$ such as that computed in \cite{NoronhaHostler:2008ju,NoronhaHostler:2012ug} or bulk viscosity \cite{Noronha-Hostler:2015qmd} can be explored in the future. 

\section{Results}

\begin{figure}[htb]
\centering
\begin{tabular}{c c}
\includegraphics[height=1.7in]{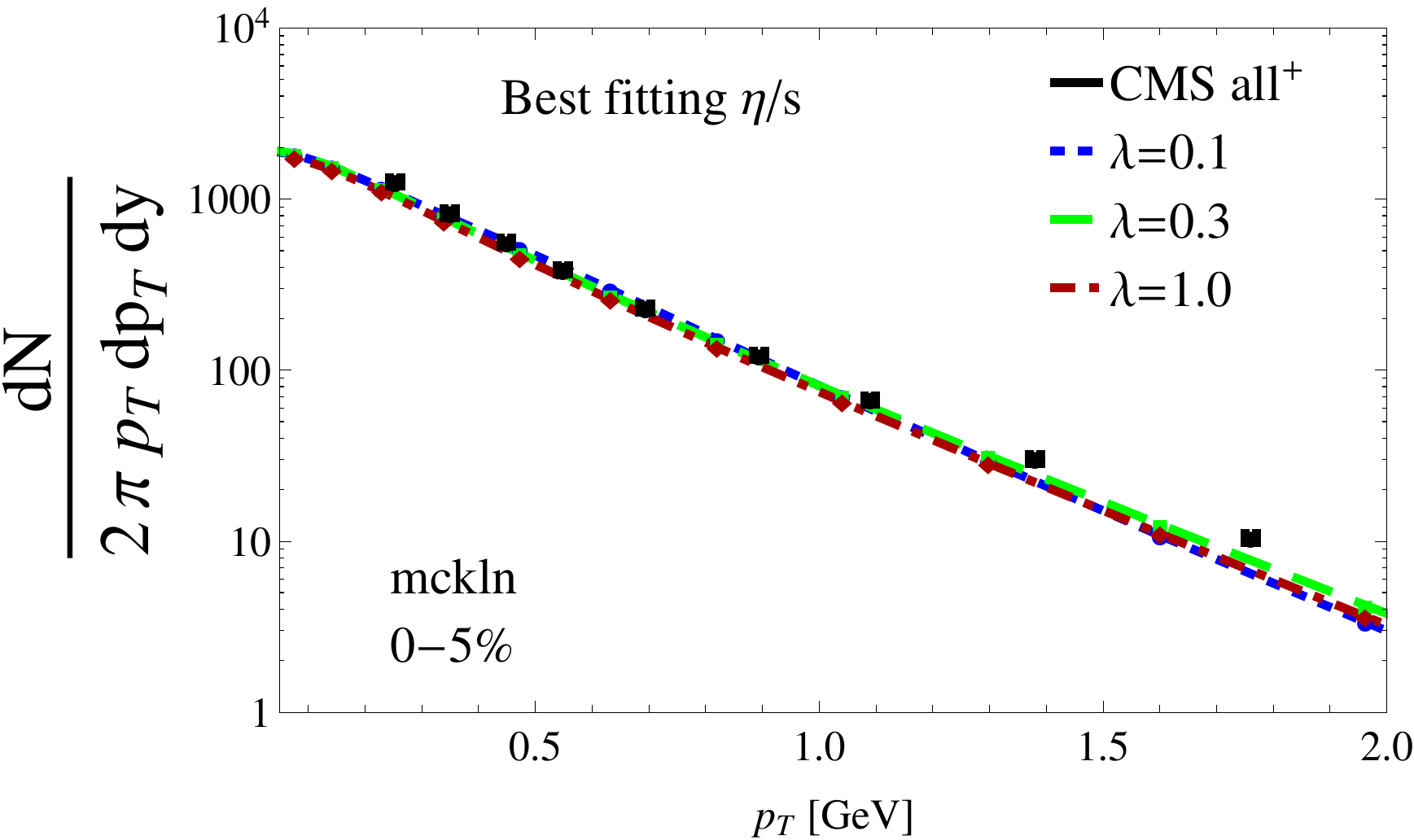} & \includegraphics[height=1.7in]{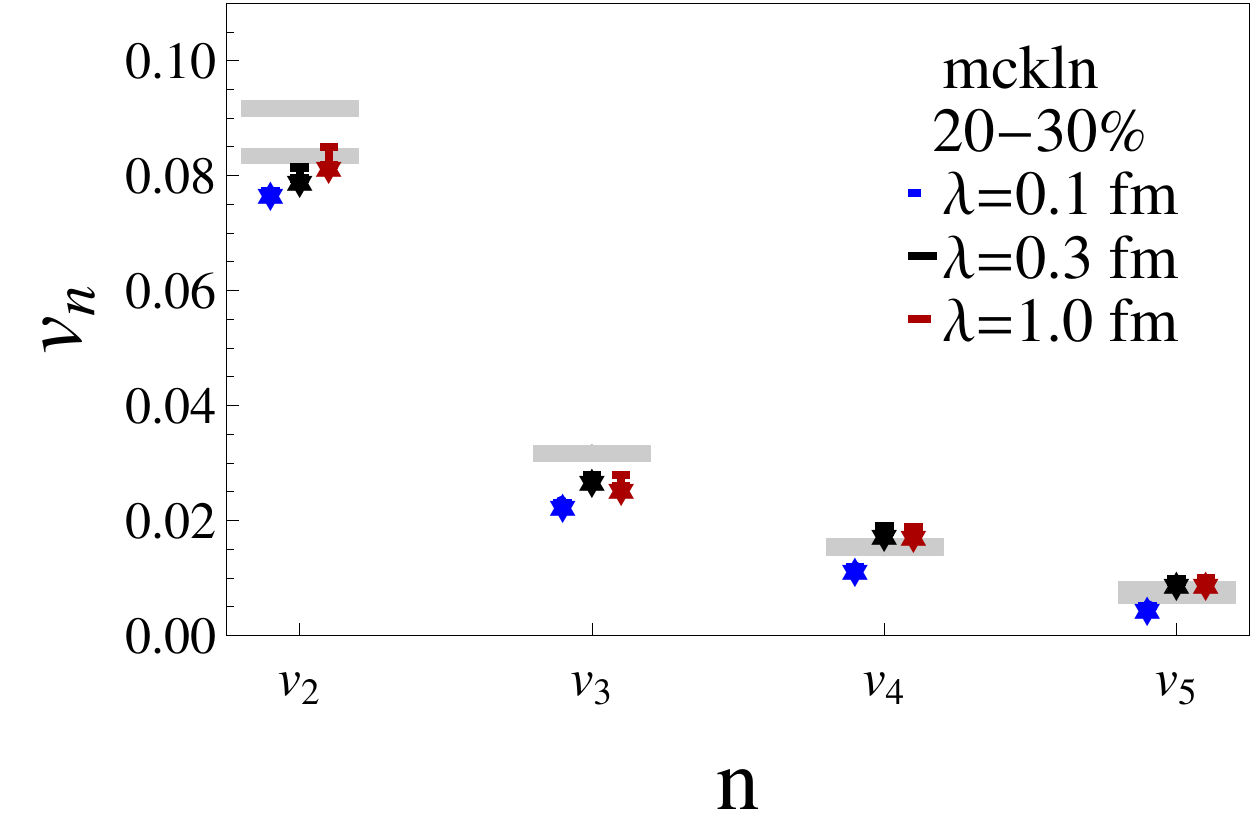}
\end{tabular}
\caption{(Color online) Effect of the smoothing scale on the spectrum of all charged hadrons (left) and integrated $v_2-v_5$ (right). Data points correspond to CMS data \cite{Chatrchyan:2012ta,Chatrchyan:2013kba}. }
\label{fig:specvn}
\end{figure}

In Fig. \ref{fig:specvn} (left) the results for the spectrum of all charged hadrons is shown for different values of the smoothing scale $\lambda$.  It is clear that the spectrum is very robust as one smooths out the energy density an entire order of magnitude from $\lambda=0.1-1.0$ fm. One should also note that $\lambda=0.1$ fm is below the microscale set by the relaxation time $\tau_{\pi}$.  In \cite{Noronha-Hostler:2015coa} when $\lambda<\tau_{\pi}$ it was shown that the Knudsen number becomes very large, which is an indication that there is not a large enough separation of scales to allow for a hydrodynamical picture.  However, as one can see in Fig.\ \ref{fig:specvn} the influence from the ``bad" Knudsen number is not seen in the spectrum, which is simply too robust.  

In Fig. \ref{fig:specvn} (right) the integrated flow harmonics $v_{2}-v_{5}$ are shown compared to CMS data \cite{Chatrchyan:2012ta,Chatrchyan:2013kba}.  Unlike for the spectrum, the effects of having ``bad" Knudsen numbers for $\lambda=0.1$ fm are more obvious in the flow harmonics.  For $v_{2}-v_{5}$, $\lambda=0.1$ fm is clearly lower than the other $\lambda$'s.  However, as long as $\lambda>\tau_{\pi}$ then there is very little difference for $v_{2}-v_{5}$ between the microscale to the mesoscale. Note that all calculations done here are for the best fitting $\eta/s$ where it was found that $\eta/s$ increases slight with $\lambda$. If one uses ideal hydrodynamics there is a stronger dependence on $\lambda$ \cite{Noronha-Hostler:2015coa}. 

Our calculations qualitatively agree with the study in Ref.\ \cite{Floerchinger:2013rya} in that the azimuthal anisotropies are much more sensitive to intermediate length scales (the mesoscopic regime in our notation) than to scales deep in the microscopic, sub-nucleonic scale regime. Once $\lambda$ is brought to the macroscopic nuclear scale regime, the eccentricities begin to change significantly, the structure in the initial conditions is lost, and nontrivial azimuthal anisotropies such as triangular flow are not produced.

\section{Conclusions}

In this paper we showed that the typical low $p_T$ experimental observables that describe collective flow are insensitive to the energy density scale between the microscale up to the mesoscale in heavy-ion collisions.  Both the particle spectrum and the integrated flow harmonics see essentially no difference as one smooths out the energy density fluctuations. This confirms previous work showing that the eccentricities alone are excellent predictors in determining the final flow harmonics. 

It is unlikely that higher temperatures reached such as those achieved at the LHC run2 could help to improve the sensitivity to the energy density scale, especially since the effect to flow harmonics is likely to be less than a $5\%$ change \cite{Noronha-Hostler:2015uye}.  Rather, the most likely collisions that may provide some indication of the energy density scale would be in small systems such as $pPb$ collisions where the $\varepsilon_{n,m}$ are significantly more affected by the smoothing scale such (in this case it was found that the eccentricities can only be smoothed out to about $\lambda=0.3-0.5$ fm \cite{Noronha-Hostler:2015coa}).  

While the known experimental flow observables do not display a strong sensitivity to sub-nucleonic fluctuations, it is possible that novel observables currently being explored could be more sensitive.  In \cite{Noronha-Hostler:2015dbi} it was found that there is about a $10\%$ non-geometrical contribution on an event-by-event basis that is washed out when one averages over an ensemble of events.  However, correlations on an event-by-event basis \cite{Bilandzic:2013kga} may preserve some of those effects.  Furthermore, in \cite{Pang:2015zrq} the effect longitudinal fluctuations are found to decorrelate elliptical and triangular flow and it would be interesting to explore the scale of those fluctuations, which is left for a future work.

We thank F.~Gardim for providing NeXus initial conditions. JNH and MG acknowledge support from the US-DOE Nuclear Science Grant No. DE-FG02-93ER40764. JN thanks FAPESP and CNPq for financial support.  

\bibliographystyle{elsarticle-num}
\bibliography{<your-bib-database>}

\end{document}